\documentclass[prl,twocolumn, aps,amssymb,footinbib,showpacs]{revtex4-1}

\usepackage{graphicx}
\usepackage{amsmath,amssymb}
\usepackage{cancel}
\usepackage{color}
\usepackage{pifont}
\usepackage{float}
\usepackage{pifont}

\begin{document}

\title{Many-body localization protected quantum state transfer}

\author{N. Y. Yao$^{1}$, C. R. Laumann$^{2}$, A. Vishwanath$^{1}$}

\affiliation{$^{1}$Department of Physics, University of California, Berkeley, CA 94720, U.S.A.}
\affiliation{$^{2}$Department of Physics, University of Washington, Seattle, WA 98195, U.S.A.}

\begin{abstract}

In thermal phases, the quantum coherence of individual degrees of freedom is rapidly lost to the environment. 
Many-body localized (MBL) phases limit the spread of this coherence and appear promising for quantum information applications. 
However, such applications require not just long coherence times but also a means to transport and manipulate  information. 
We demonstrate that this can be done  in a one dimensional model of interacting spins at infinite temperature. 
Our protocol utilizes protected qubits which emerge at the boundary between topological and trivial phases. 
State transfer occurs via dynamic shifts of this boundary and is shown to preserve quantum information.  
As an example, we discuss the implementation of a universal, two-qubit gate based upon MBL-protected quantum state transfer. 
\end{abstract}


\maketitle


%

The faithful storage and transmission of quantum information is the basic building block for applications ranging from information processing to communication and metrology \cite{Bennett93, Gisin02,Lloyd93}. 
In a traditional architecture, quantum bits connect via channels that coherently shuttle information between remote nodes \cite{Kimble08}. 
Constructing such a channel from an interacting many-body system at high temperature is generally considered impossible; once quantum information disperses into the system, it rapidly decoheres due to scattering with thermal excitations. 
To avoid this, typical quantum channels,  such as mechanical resonators \cite{McGee13,Palomaki13}, optical photons \cite{Blinov04, Duan04}, superconducting strip-lines \cite{Sillanpaa07,Majer07}, and spin chains \cite{Bose03, Petrosyan10, Christandl04, Burgarth07, Yung05, Difranco08, Kay07, Feldman10, Clark05, Venuti07, Gualdi08, Paternostro05, Tsomokos07, Wojcik05, Banchi10,Yao11}, are either specially tailored few-body systems or operate at ultra-low temperatures in order to freeze out parasitic degrees of freedom.

Recently, an alternate strategy to deal with thermal excitations has emerged: localization \cite{Anderson58,Fleishman80,Altshuler97,Basko06,Gornyi05}.  The introduction of sufficiently strong quenched disorder leads to a \emph{many-body localized} (MBL) phase \cite{Anderson58,Fleishman80,Altshuler97,Basko06,Gornyi05,Burin94,Burin98,Burin98b,Burin06,Oganesyan07,Pal10,Znidaric08,Monthus10,Bardarson12,Vosk12,Iyer13,Serbyn13,Huse13,Serbyn13b,Huse13b,Pekker13,Vasseur15,Potter15,Potter15b,Vosk13,Bahri13,Chandran13,Bauer13,Swingle13,Schiulaz13,Serbyn14,Ros15,Chandran15,Agarwal15,Gopalakrishnan15,BarLev15}  with  effective Hamiltonian, 
\begin{align}
	\label{eq:hamlbits}
	H = \sum_i  \mathcal{J}_i  \tau^z_i + \sum_{ij}  \mathcal{J}_{ij}   \tau^z_i \tau^z_j + \sum_{ijk}  \mathcal{J}_{ijk}   \tau^z_i \tau^z_j \tau^z_k + \cdots
\end{align}
where $\tau^z_i$ are local integrals of motion and $\mathcal{J}_{ij\cdots}$  decay exponentially \cite{Serbyn13b,Huse13b,Ros15,Chandran15}. In general, these  $\tau^z_i$ are locally dressed versions of the underlying physical degrees of freedom.
From Eqn.~\eqref{eq:hamlbits}, it is clear that their $z$-polarization never decays (i.e. $T_1 \rightarrow \infty$). However, superpositions dephase rapidly due to the Ising-type interactions with neighbors  (i.e. $T_2^* \sim 1/\mathcal{J}$). In principle, for each $\tau^z_i$ this dephasing can be canceled by a local spin-echo  (i.e. $T_2 \rightarrow \infty$) \cite{Huse13b,Serbyn14}. 
In practice, there are three significant challenges: 1) there is only a finite random overlap between the  physically addressable degrees of freedom and the $\tau^z_i$, 2) quantum information is only protected in the presence of continuous spin-echo, 3) it is unclear how to achieve robust quantum state transfer and logic gates. 

\begin{figure}
	\centering
		\includegraphics[width=3.4in]{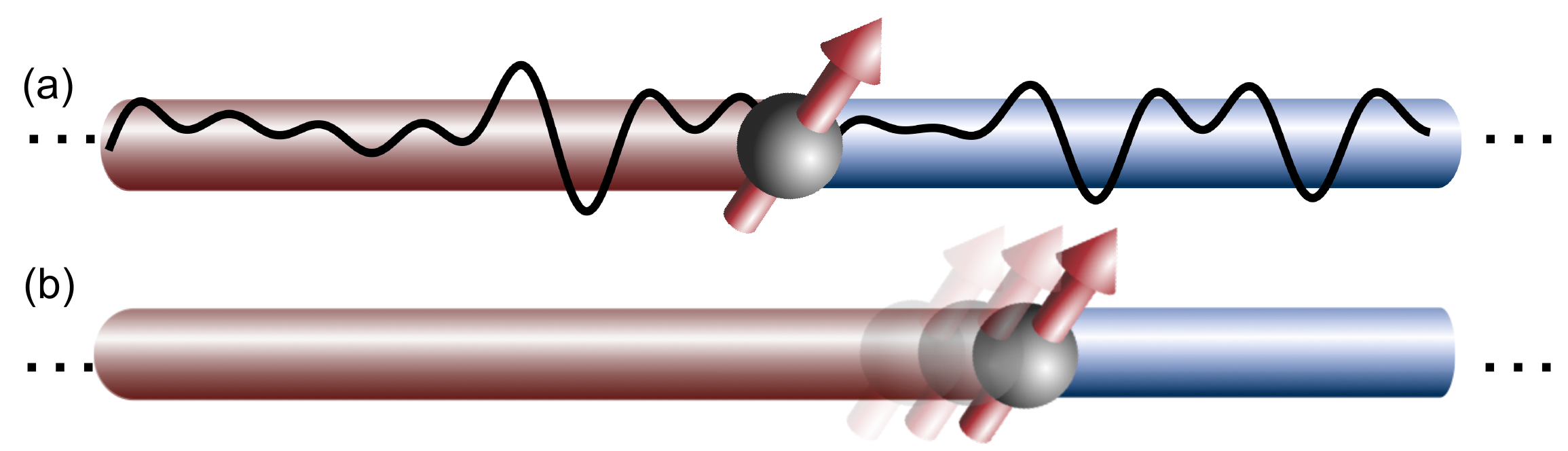}
	\caption{%
	(a) Schematic representation of the boundary between a trivial and SPT insulator. A single qubit is bound at the interface. Strong disorder (black line) leads to many-body localization in both phases.
	(b) Shuttling of the edge qubit occurs via shifts in the trivial-SPT boundary.
}
	\label{fig:fig1_v1_schem}
\end{figure}

In this Letter, we demonstrate the storage and transmission of quantum information in an interacting, many-body spin chain at infinite temperature (i.e. finite energy density) \cite{finitedensity}.
Our approach requires neither initialization nor continuous spin-echo. 
Rather, we exploit the boundary qubit  at the interface between a many-body localized trivial and symmetry protected topological (SPT) phase (Fig.~1) \cite{Huse13b,Bahri13,Chandran13,Potter15}. Symmetry prevents the operators corresponding to this degree of freedom from appearing in the effective Hamiltonian [Eqn.~\eqref{eq:hamlbits}], implying that  $T_2^* \rightarrow \infty$ even in the absence of spin-echo. Our main result is summarized in Table~I (bottom row);  the $\mathcal{O}(1)$ local spectral gap of the MBL phase is sufficient to protect static storage (Fig.~2a), dynamic manipulations (Fig.~2b), and quantum gates (Fig.~2c) between boundary qubits; we emphasize that our approach does not rely upon, nor work in, the exponentially slow, strictly adiabatic limit \cite{Khemani15}.

%

%
%


\emph{Protocols and model}---In one dimension, the canonical example of an SPT phase is provided by the Haldane phase, whose boundary binds a single qubit (Fig.~1a) \cite{Haldane83,Affleck87}. But what exactly does this mean? In one sense, this means that  coherent quantum information may be stored and retrieved from the edge qubit. Let us define a ``static'' protocol to sharpen this: measure the $\alpha$-component, $\Sigma^\alpha$, of the qubit to initialize it,  wait for a time $\tau$, and then measure $\Sigma^\alpha$ again. 
At zero temperature,  the measurements coincide and there is neither loss of polarization nor coherence.

Being bound to the edge  also implies that the qubit moves when the boundary shifts (Fig.~1b). But, does this motion decohere the quantum information? Let us define a corresponding ``dynamic'' protocol: measure $\Sigma^\alpha$, move the boundary of the SPT phase (e.g. by ramping local fields and converting a region into the opposite phase) on a time-scale $\tau_R$, hold for a time $\tau$, ramp back and measure. At zero temperature, the measurements again coincide when $\tau_R \gg 1/\Delta^2$, where $\Delta$ is the bulk gap, as required by the quantum adiabatic theorem (Table~I, top row).

  \begin{table}
\centering 
 \caption{MBL protected storage (static) and shuttling (dynamic)} 
\begin{tabular}{c |c |c}
 \hline\hline &Static & Dynamic  \\
\hline 
Zero $T$, clean  & $\checkmark$ & $\checkmark$   \\
Finite $T$, clean, non-interacting  & $\checkmark$& \ding{53}   \\
Finite $T$, clean, interacting &  \ding{53} &  \ding{53}    \\ 
Finite $T$, disordered, interacting & $\checkmark$  & $\checkmark$  \\ \hline 

 \end{tabular} \label{table:critdim} 
\end{table}

The introduction of finite temperature invalidates the above lore. Surprisingly, a distinction emerges between the `static' and `dynamic'  protocols (Table~I). 
Consider the following microscopic spin Hamiltonian \cite{Bahri13},
\begin{align}
	\label{eq:hamgeneral}
	H = \sum_i \lambda_i  \sigma^z_{i-1} \sigma^x_i \sigma^z_{i+1} + \sum_i h_i \sigma^x_i + \sum_i V_i \sigma^x_i \sigma^x_{i+1}
\end{align}
where $\sigma^\alpha$ are  Pauli operators. This model has a $\mathbb{Z}_2 \times \mathbb{Z}_2$  symmetry corresponding to products of $\sigma^x$ on the even and odd sites. The symmetry protects the Haldane phase realized when $\lambda$ is the dominant coupling. A trivial insulator arises when $h \gg \lambda,V$ and thus a trivial-SPT boundary can be realized (Fig.~1a) by spatially varying the ratio $\lambda/h$. In the simplest case  when $V=0$ one can write down an explicit expression for the edge qubit. Setting $h_i=0$ on the SPT side ($i\geq 1$) and $\lambda_i=0$ on the trivial side ($i\leq -1$), the Hamiltonian is a sum of commuting terms. One finds that the following Pauli operators at the boundary commute with the Hamiltonian,  $\Sigma^x = \sigma^x_0 \sigma^z_1$, $\Sigma^y = \sigma^y_0 \sigma^z_1$, $\Sigma^z = \sigma^z_0$, but are odd under the $\mathbb{Z}_2 \times \mathbb{Z}_2$ symmetry and hence do not appear in $H$. The Pauli algebra guarantees that they generate a two fold degenerate edge state.  For a more generic Hamiltonian,  the edge states are dressed by exponentially localized tails.

 \begin{figure}
	\centering
		\includegraphics[width=3.4in]{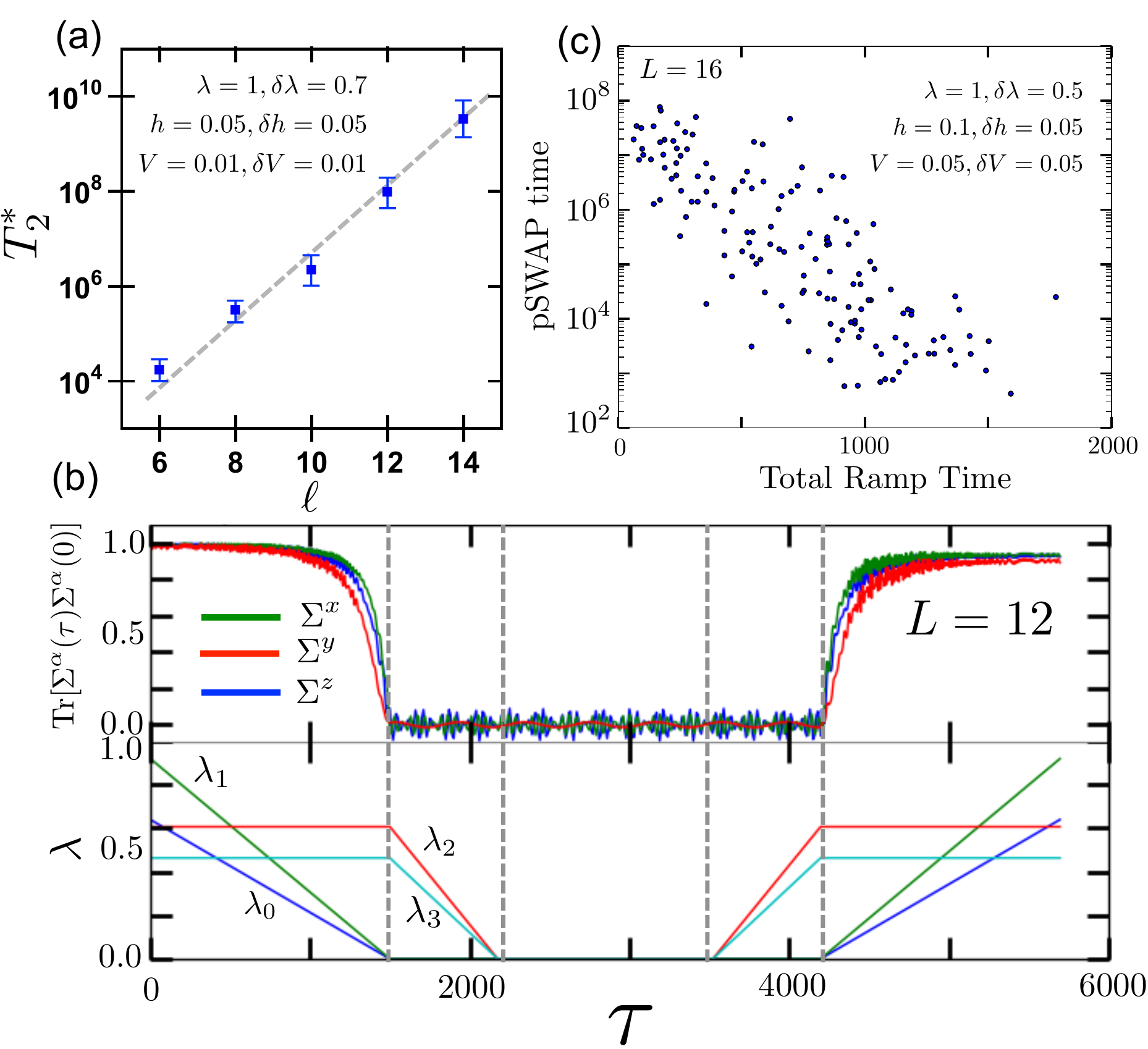}
	\caption{%
	Exact diagonalization and numerical integration of Hamiltonian evolution in the interacting, disordered, infinite temperature case. All numerics are performed using a box distribution with mean and width shown in the figure. 
	(a) 
	Depicts the static coherence time of the edge qubit, which grows exponentially with system size. The total system size is $L=\ell+2$ with a trivial insulator ($\lambda = 0$) occupying the first two sites followed by an SPT region of length $\ell$.
	(b) Fidelity of the dynamic protocol. The boundary qubit is successively shuttled inwards by a total of four lattice sites with the ramp profile shown in the bottom panel.  (c) 
	 Shows $145$ disorder realizations of the dynamic protocol for $L=16$. In each realization, the boundary qubit is shuttled by a total of eight lattice sites in four successive steps. The pSWAP gate time improves exponentially while the adiabatic ramp time is governed by the $\mathcal{O}$(1) local spectral gap.   }
	\label{fig:fig1_v1_schem}
\end{figure}

To analyze the protocols, we switch to a convenient representation of the Hamiltonian in terms of Majorana fermions:
\begin{align}
	\label{eq:hammajorana}
	H =  \sum_i i  \lambda_i \gamma^2_{i-1} \gamma^1_{i+1} +  \sum_i  i h_i \gamma^1_i \gamma^2_i - \sum_i V_i \gamma^1_i \gamma^2_i \gamma^1_{i+1} \gamma^2_{i+1}.
\end{align}
Here,  each spin is represented by a pair of real fermions $\gamma^1$ and $\gamma^2$, with the Pauli algebra given by, $\sigma^x_i = i \gamma^1_i \gamma^2_i$, $\sigma^y_i =  \prod_{j<i} \left( i \gamma^1_j \gamma^2_j \right) \gamma^2_i$, $\sigma^z_i =  \prod_{j<i} \left( i \gamma^1_j \gamma^2_j \right) \gamma^1_i$.

\emph{Finite temperature, clean}---We begin the finite temperature analysis with $V=0$ in the clean case (uniform couplings), where the Hamiltonian is quadratic and has a simple pictorial representation (Fig.~3a). This clarifies the nature of the  $\mathbb{Z}_2 \times \mathbb{Z}_2$ symmetry: the fermions hop on two chains, each of whose parity is conserved. In the fermionic language, the edge qubit operators are given by the zero modes ($\gamma^1_0$, $\gamma^1_1$ triangles, Fig.~3a) along with Jordan-Wigner strings that lie entirely in the trivial phase \cite{parity}, 
\begin{align}
	\label{eq:JQ}
	\Sigma^x &= \sigma^x_0 \sigma^z_1  = \prod_{j<0} \left( i \gamma^1_j \gamma^2_j \right) \gamma^1_1 \nonumber \\
	\Sigma^y &= \sigma^y_0 \sigma^z_1  = -i \gamma^1_0  \gamma^1_1 \nonumber \\ 
	\Sigma^z &= \sigma^z_0 = \prod_{j<0} \left( i \gamma^1_j \gamma^2_j \right) \gamma^1_0.
\end{align}
These strings reflect the SPT nature of the qubit; they attach to all local operators odd under the $\mathbb{Z}_2 \times \mathbb{Z}_2$ symmetry.

In the static protocol, both the zero modes and the Wigner string are conserved and thus   coherence persists  even at finite temperature (Table~I, second row). 
The dynamic protocol is more subtle. 
To shift the boundary, one ramps up $\lambda_0(t), \lambda_{-1}(t)$ as shown in Fig.~3c. At all times during this ramp, there is a unique zero mode with an $\mathcal{O}(1)$ gap to the bulk for each chain; thus, if each ramp proceeds slower than this gap scale, the zero mode simply shifts and returns with no phase accumulation. 
However, the last four fermionic operators of the Wigner string ($\gamma^1_{-2},\gamma^2_{-2},\gamma^1_{-1},\gamma^2_{-1}$, Fig.~3a) are carried across the phase boundary and necessarily move into modes above the gap.
At $T=0$, there are no fermionic excitations in the system that can be carried across with these modes; however, at finite temperature, any excitation carried across the boundary disperses into the bulk of the SPT phase. Thus, 
the qubit's coherence is preserved through the dynamic protocol only at $T=0$ in the clean system.

To confirm this picture, we consider the trace fidelity of the edge qubit,
\begin{align}
	\label{eq:hamgeneral}
	F^\alpha(t) = \frac{1}{Z} \text{Tr}\left [e^{-\beta H} \Sigma^\alpha(t) \Sigma^\alpha(0) \right],
\end{align}
as a function of time, where $Z$ is the partition function and $\beta = 1/k_BT$. This fidelity is precisely the expected correlation of the measurements in the static and dynamic protocols. We numerically simulate the dynamic protocol at infinite temperature on finite size systems of length $L$ with the boundary at $L/2$. For large systems, we expect the dynamic protocol to fail (Fig.~3b), while for sufficiently small systems, it is possible for the ramp rate to resolve the mini-gaps $\sim 1/L$ within the bulk bands. 
Indeed, for fixed  $\tau_R$, the qubit's coherence is lost as $L$ increases (Fig.~3b inset) as neither $\Sigma^x$ nor $\Sigma^z$ recover (Table~I, second row). The persistence of $\Sigma^y$ reflects its lack of a Wigner string and further confirms our above analysis.

In the interacting case, $V\neq0$, the situation is much worse: both the static and dynamic protocols fail at finite temperature (Table~I, third row). Indeed, scattering by bulk excitations leads to a static coherence time $T_2^* \sim \lambda/V^2$ \cite{SuppInfo}. 
Until recently, this would be the end of the story. However, it is now believed that strong disorder, leading to many-body localization can protect SPT order even at finite temperature \cite{Huse13b,Bahri13,Chandran13,Potter15}. 

\begin{figure}
	\centering
		\includegraphics[width=3.4in]{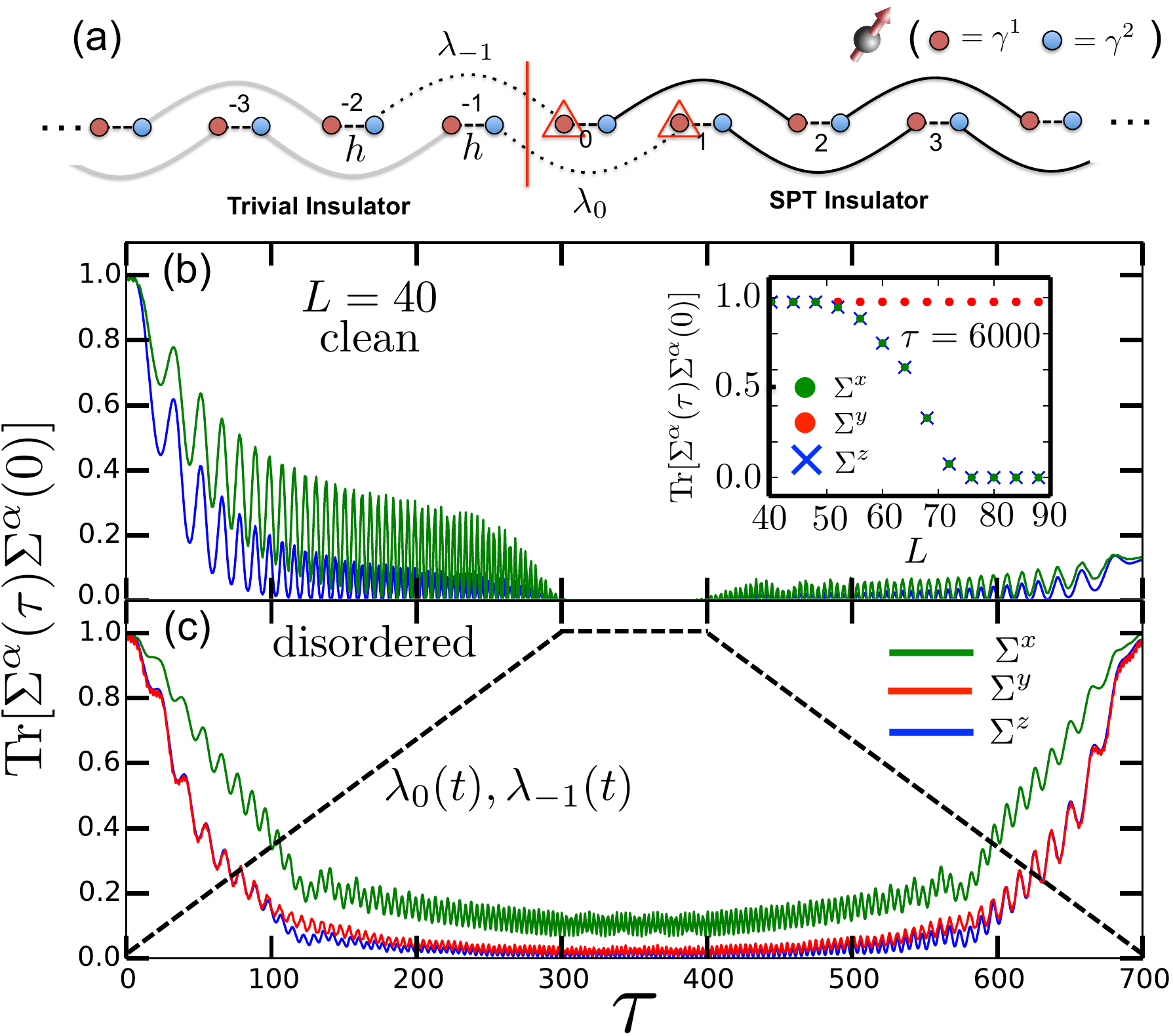}
	\caption{%
	(a) The $V=0$ limit in which $H$ reduces to a pair of decoupled fermion chains. The SPT insulator (on the right, $i \ge 0$) has $\lambda$ (black line) $\gg$ h (dotted line) while the trivial insulator (on the left, $i < 0$) has $\lambda$ (gray line) $\ll$ h (dotted line). Ramping the boundary terms $\lambda_{0,-1}$, shifts the interface and hence the location of the edge qubit. 
	(b) Trace fidelity of the dynamic protocol in the infinite temperature, non-interacting, clean case for a chain of length  $L=40$ with the boundary at $L/2$. On the SPT side, $h = 0.1$, $\lambda = 1$  while on the trivial side, $h = 0.1$, $\lambda = 0.01$. The couplings $\lambda_0$ and $\lambda_{-1}$  ramp linearly from $0.01$ to $1$ as a function of time [dotted line in (c)].  (inset) For $L=40$, a ramp time of $\tau_R \sim 10^4$ is slow enough to resolve the mini-gaps $\sim 1/L$ within the band; thus, the dynamic protocol exhibits a nearly perfect fidelity. However, this is a finite size effect and increasing the system size immediately leads to a failure of the protocol. (c) Same as (b) but with strong disorder, $\delta \lambda =0.7 \lambda$, $\delta h =h$. Localization of the Wigner strings enables a recovery fidelity $>99\%$ with the same ramp time. }
	\label{fig:fig1_v1_schem}
\end{figure}

\emph{Finite temperature, disordered}---The localization of bulk excitations suppresses  scattering and enables the static coherence of the edge qubit to persist in this MBL SPT phase. More specifically, in the trivial insulator, the local integrals of motion are given by $\tau^z_i = U \sigma^x_i U^\dagger$, while in the SPT insulator,  $\tau^z_i = U \sigma^z_{i-1} \sigma^x_i \sigma^z_{i+1} U^\dagger$,
 where $U$ is a local $\mathbb{Z}_2 \times \mathbb{Z}_2$ symmetric unitary  that diagonalizes the Hamiltonian. These symmetry-even local operators appear directly in Eqn.~\eqref{eq:hamlbits}. 
%
As previously discussed, unlike the bulk $\tau^z_i$, the boundary qubit, $\tau^z_b = U  \Sigma^z U^\dagger$ is symmetry odd and therefore cannot appear in  Eqn.~\eqref{eq:hamlbits}. Instead, the boundary qubit can only be dephased by interactions involving other edge modes. For an SPT region of length $\ell$, this implies $T_2^* \sim e^\ell$, as confirmed numerically in Fig.~2a.

But, can localization protect coherence during dynamical protocols? 
The non-interacting case already provides a tantalizing hint that this may be. Figure~3c shows the successful recovery of quantum information in a disordered system at infinite temperature. We note that the ramp time used  is identical to that of the failed attempt in the clean system (Fig.~3b). 
Here, localization plays a crucial, but different role than in the static case. It prevents the dispersion of the piece of the  Wigner string carried across the boundary, limiting its spread to a localization length $\xi$ independent of system size. 
Thus, so long as the ramp time is adiabatic with respect to the local gap, $\tau_R \sim 1/\lambda \xi$, the string  returns to its original position following the protocol.

In the interacting case, one expects that the adiabaticity condition is  modified by at most a factor $2^\xi \sim \mathcal{O}(1)$. Numerics confirm this expectation: The top panel of Fig.~2b depicts the fidelity of a successful dynamic protocol at infinite temperature, with both strong interactions and disorder. Two successive ramps (ramp profile shown in Fig.~2b bottom) move the edge qubit a total of four lattice sites before holding and returning; the system size is $L=12$ with the trivial phase initially occupying the first two sites. 
The example shown is typical among $\sim10^3$ samples; failure (less than $90\%$ recover fidelity) arises for less than $3\%$ of the total disorder realizations and results mainly from two causes: 1) when there are additional, unexpected, trivial-SPT boundaries and 2) when the localization length is significantly larger than a few lattice sites.

\emph{MBL protected quantum gates}---Between the static and dynamic protocols, we now have robust control over a single edge qubit in the MBL SPT phase. This raises the question of whether one can perform a non-trivial gate between two such qubits. For a finite SPT region of length $\ell$, the two edge qubits interact with each other through their exponentially localized tails. In the clean, non-interacting case,  this leads to coherent  oscillations on a time scale, $\tau_0 = \frac{\pi}{h} (\frac{\lambda}{h})^{\ell/2-1}$ \cite{SuppInfo}.

The nature of these  oscillations is apparent from the fermionic  Hamiltonian in Eqn.~\eqref{eq:hammajorana} at $V=0$. At time $\tau_0$, the zero modes at each end of the SPT region have swapped position. In the language of the edge qubits, 
\begin{align}
	\label{eq:JQ}
	\Sigma^z_L (t) &= \cos(\phi_A t) \Sigma^z_L (0) + \sin(\phi_A t) i P \Sigma^x_R (0) \nonumber \\
	\Sigma^x_L (t) &=\cos(\phi_B t) \Sigma^x_L (0) - \sin(\phi_B t) i P \Sigma^z_R (0)  \nonumber \\ 
	\Sigma^z_R (t) &= - \sin(\phi_B t) i P \Sigma^x_L (0) + \cos(\phi_B t) \Sigma^z_R (0)  \nonumber \\
	\Sigma^x_R (t) &= \sin(\phi_A t) i P \Sigma^z_L (0) + \cos(\phi_A t) \Sigma^x_R (0), 
\end{align}
where $L,R$ index the left and right qubits, $\phi_A$, $\phi_B$ are the oscillation frequencies ($\phi = 2\pi/\tau_0$ for the clean case), and $P = \prod_i \sigma^x_i$ is the parity operator for the SPT region.

\begin{figure}
\center
\includegraphics[width=3.0in]{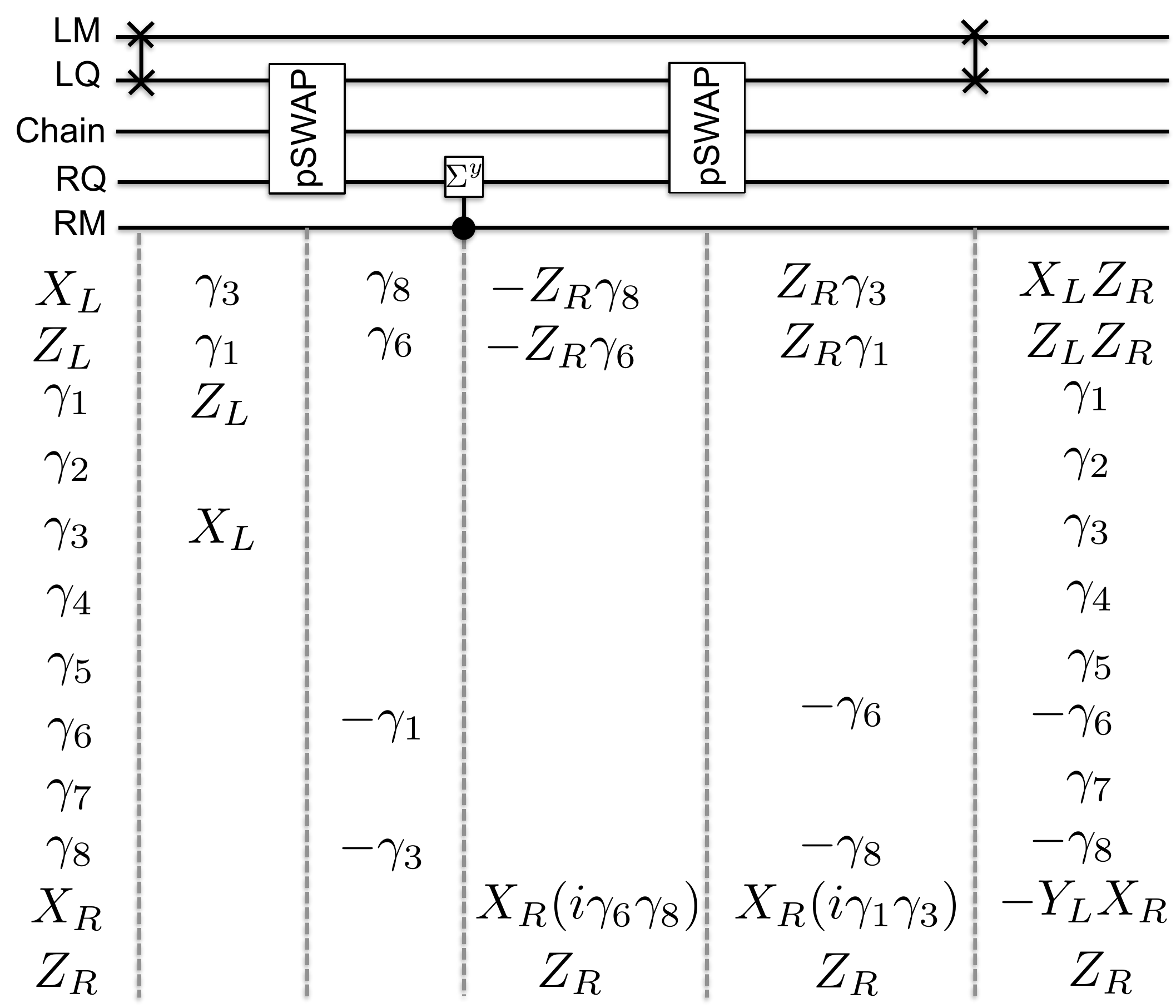}
\caption{Quantum circuit demonstrating a universal MBL-protected gate between ancillae qubits $LM$ and $RM$.
The Heisenberg evolution of all operators of an $L=6$ SPT region is shown following each gate. 
At the end of the evolution, a universal controlled-$Y$ gate has been implemented between $LM$ (control) and $RM$ (target). 
}
\label{fig:arch1}
\end{figure}

At time $\tau_0$, Eqn.~\eqref{eq:JQ}  produces a non-trivial gate between the edge qubits, which we term: pSWAP. It corresponds to a SWAP gate between the edge qubits entangled with the parity of the SPT region. 
This entanglement is rather benign since $P^2 = \mathbb{I}$, implying that a second pSWAP cancels the unwanted entanglement. 
Unfortunately, the naive pSWAP gate suffers from two issues: 1) it does not directly lead to universal quantum computation and 2) it is exponentially slow in the separation of the edge qubits. 

The first issue is easily remedied by placing an ancilla qubit next to each SPT edge qubit. With full control over this quantum register, we can perform an MBL-protected universal quantum gate, as shown in the quantum circuit of Fig.~4. The explicit Heisenberg evolution of the operator basis is indicated below the circuit ($L=6$). From this, one can deduce that a universal controlled-Y gate occurs between the two ancillae. 

The second issue can be remedied by using the dynamic protocol for quantum state transfer.  By shuttling the edge qubits closer to one another, the pSWAP gate time is exponentially improved. 
The coherence time of the SPT edge qubits $\sim e^\ell$. Meanwhile, each shuttling step takes time $\sim 1/h^2$; following $m$ steps, the pSWAP gate occurs in a time $\sim e^{\ell-m}$ giving an exponential enhancement between gate and coherence times.
This is explicitly shown in Fig.~2c; we simulate $145$ disorder realizations on an interacting $L=16$ chain at infinite temperature. 
We shuttle the left SPT qubit from $2$ to $8$ sites. At  each  separation, we extract the pSWAP time and scatter this against the total ramp time. Each ramp  is optimized  to ensure a final fidelity $F > 90\%$.
 In the disordered case, since $\phi_A \neq \phi_B$ in general, one must match the phase accumulated by $\Sigma^x$ and $\Sigma^z$ [Eqn.~\eqref{eq:JQ}]. This is accomplished by adjusting the hold time between each ramp.
 
\emph{Conclusion}---The binding of a qubit to the interface between trivial and SPT insulators has enraptured a generation of theoretical physicists. At zero temperature, shifting the position of this interface coherently drags the qubit along, analogous to the shuttling of Majorana fermion zero modes in topological superconducting wires \cite{Alicea11,Carmele15}. In clean systems at finite temperature, neither storage nor shuttling of the SPT qubit works. We have demonstrated that strong disorder, leading to many-body localization, protects  qubit coherence in both ``static'' and ``dynamic'' protocols. Localization plays two roles: it prevents scattering from thermal excitations in the bulk and also prevents the ``Cherenkov'' emission of quantum information during transport.  This enables the coherent control of qubits in disordered, strongly interacting systems at high temperature. Furthermore, our construction of the pSWAP gate demonstrates that these qubits may be wired up and universal gates performed between them. Looking forward, it would be interesting to consider a wire network of disordered SPT insulators as a platform architecture \cite{Alicea11}.   
The most natural system where our approach offers a competitive advantage is one in which Hamiltonian engineering is possible but cooling into fiducial initial states is difficult. One intriguing possibility is in driven systems where ``stroboscopic'' SPT phases can emerge upon periodic modulation of  interactions \cite{Iadecola15}. In these cases, many-body localization is, by itself, unable to provide coherent local degrees of freedom, thus motivating the need for an MBL SPT qubit. 

It is a pleasure to gratefully acknowledge the insights of and discussions with D. Abanin, Y. Bahri, S. Choi, Y. Gannot, S. Gopalakrishnan, D. Huse, M. Lukin,  M. Knap, R. Nandkishore,  V. Oganesyan, S. Parameswaran, A. Potter, R. Vasseur. N. Y. Y. acknowledges support from the Miller Institute for Basic Research in Science. A.V. acknowledges support from AFOSR MURI grant FA9550-14-1-0035.


\begin{thebibliography}{99}

\bibitem{Bennett93} C. H. Bennett \emph{et al}, Phys. Rev. Lett. {\bf70}, 1895 (1993).

\bibitem{Gisin02} N. Gisin \emph{et al}, Rev. Mod. Phys. {\bf74}, 145 (2002).

\bibitem{Lloyd93} S. Lloyd, Science {\bf261}, 5128 (1993).

\bibitem{Kimble08} H. J. Kimble, Nature {\bf453}, 1023 (2008).



\bibitem{McGee13} S. A. McGee, D. Meiser, C. A. Regal, K. W. Lehnert, M. J. Holland, Phys. Rev. A {\bf87}, 053818  (2013).

\bibitem{Palomaki13} T. A. Palomaki,	J. W. Harlow,	J. D. Teufel,	R. W. Simmonds,	 K. W. Lehnert, Nature {\bf495}, 210 (2013).




\bibitem{Blinov04} B. B. Blinov \emph{et al}, Nature {\bf428}, 153-157 (2004).

\bibitem{Duan04} L. M. Duan \emph{et al}, Quant. Inf. and Comp. {\bf4}, 165-173 (2004).


\bibitem{Sillanpaa07} M. A. Sillanpaa, J. I. Park, R. W. Simmonds, Nature {\bf449}, 438 (2007).

\bibitem{Majer07} J. Majer \emph{et al}, Nature {\bf449}, 443 (2007).




\bibitem{Bose03} S. Bose, Phys. Rev. Lett. {\bf91}, 207901 (2003).\

\bibitem{Christandl04} M. Christandl \emph{et al}, Phys. Rev. Lett. {\bf92}, 187902 (2004).

\bibitem{Wojcik05} A. Wojcik \emph{et al}, Phys. Rev. A {\bf71}, 034303  (2005).

\bibitem{Difranco08} C. Di Franco \emph{et al}, Phys. Rev. Lett. {\bf101}, 230502  (2008).

\bibitem{Paternostro05} M. Paternostro \emph{et al}, Phys. Rev. A {\bf71}, 042311  (2005).

\bibitem{Kay07} A. Kay, Phys. Rev. Lett. {\bf98}, 010501 (2007).

\bibitem{Venuti07} L. Campos Venuti  \emph{et al}, Phys. Rev. A {\bf76}, 052328  (2007).

\bibitem{Gualdi08} G. Gualdi  \emph{et al}, Phys. Rev. A {\bf78}, 022325  (2008).

\bibitem{Feldman10} E. B. Fel'dman \emph{et al}, Phys. Rev. A {\bf82}, 022332  (2010).

\bibitem{Petrosyan10} D. Petrosyan \emph{et al}, Phys. Rev. A {\bf81}, 042307 (2010).

\bibitem{Burgarth07} D. Burgarth \emph{et al}, Phys. Rev. A {\bf75}, 062327 (2007).

\bibitem{Yung05} M. H. Yung and S. Bose, Phys. Rev. A {\bf71}, 032310 (2005).

\bibitem{Clark05} S. R. Clark \emph{et al}, New J. Phys. {\bf7}, 124 (2005).

\bibitem{Tsomokos07} D. I. Tsomokos \emph{et al}, New J. Phys. {\bf9}, 79 (2007).

\bibitem{Banchi10} L. Banchi \emph{et al}, Phys. Rev. A {\bf82}, 052321  (2010); L.
Banchi \emph{et al}, arXiv:1012.1572v1.

\bibitem{Yao11} N. Y. Yao \emph{et al}, Phys. Rev. Lett. {\bf106}, 040505  (2011).


\bibitem{Anderson58} P. W. Anderson, {\it Phys. Rev. } {\bf109}, 1492 (1958).

\bibitem{Fleishman80} L. Fleishman and P. W. Anderson, {\it Phys. Rev. B} {\bf21}, 2366 (1980).

\bibitem{Altshuler97} B. L. Altshuler, Y. Gefen, A. Kamenev, and L. S. Levitov, {\it Phys. Rev. Lett.} {\bf78}, 2803 (1997).



\bibitem{Basko06} D. Basko, I. Aleiner, and B. Altshuler,  {\it  Ann. Phys. } {\bf321}, 1126 (2006).

\bibitem{Gornyi05} I. V. Gornyi, A. D. Mirlin, and D. G. Polyakov, {\it Phys. Rev. Lett.} {\bf95}, 206603 (2005).



\bibitem{Ros15}  V. Ros, M. Mueller, and A. Scardicchio, {\it Nucl. Phys. B} {\bf891}, 420 (2015).





\bibitem{Chandran15}  A. Chandran, I. H. Kim, G. Vidal, D. A. Abanin, {\it Phys. Rev. B} {\bf91}, 085425 (2015).


\bibitem{Serbyn13b}  M. Serbyn, Z. Papic, and D. A. Abanin, {\it Phys. Rev. Lett.} {\bf111}, 127201 (2013).

\bibitem{Huse13b}  D. A. Huse, R. Nandkishore, V. Oganesyan, A. Pal, and
S. L. Sondhi, arXiv:1305.6598 (2013).







\bibitem{Bahri13} Y. Bahri, R. Vosk, E. Altman, and A. Vishwanath,
arXiv:1307.4092 (2013)

\bibitem{Chandran13} A. Chandran, V. Khemani, C. R. Laumann, and S. L. Sondhi, arXiv:1310.1096 (2013)

\bibitem{Potter15} A. C. Potter, A. Vishwanath, arXiv:1506.00592 (2015)

\bibitem{Serbyn14}  M. Serbyn \emph{et al}, {\it Phys. Rev. Lett.} {\bf113}, 147204 (2014).



\bibitem{Burin94} A. L. Burin and Y. Kagan,  {\it J. Exp. Theor. Phys.} {\bf106}, 633-637 (1994).

\bibitem{Burin98} A. L. Burin, Y. Kagan, L. A. Maksimov, I. Y. Polishchuk,  {\it Phys. Rev. Lett.} {\bf80}, 2945 (1998).

\bibitem{Burin98b} A. L. Burin, D. Natelson, D. D. Osheroff, Y. Kagan,  {\it Tunneling Systems in Amorphous and Crystalline Solids}, Springer Verlag, 223-316 (1998).


\bibitem{Burin06} A. L. Burin, arXiv:cond-mat/0611387 (2006)

\bibitem{Oganesyan07}V. Oganesyan and D. A. Huse,  {\it Phys. Rev. B} {\bf75}, 155111 (2007).

\bibitem{Pal10} A. Pal and D. A. Huse,  {\it Phys. Rev. B} {\bf82}, 174411 (2010).

\bibitem{Znidaric08} M. Znidaric, T. Prosen, and P. Prelovsek,   {\it Phys. Rev. B} {\bf77}, 064426 (2008).

\bibitem{Monthus10} C. Monthus and T. Garel,  {\it Phys. Rev. B} {\bf81}, 134202 (2010).

\bibitem{Bardarson12} J. H. Bardarson, F. Pollmann, and J. E. Moore, {\it Phys. Rev. Lett.} {\bf109}, 017202 (2012).

\bibitem{Vosk12} R. Vosk and E. Altman,  {\it Phys. Rev. Lett.} {\bf110}, 067204 (2013).

\bibitem{Iyer13} S. Iyer, V. Oganesyan, G. Refael, and D. A. Huse, {\it Phys. Rev. B} {\bf87}, 134202 (2013).


\bibitem{Huse13}  D. A. Huse and V. Oganesyan, arXiv:1305.4915  (2013)


\bibitem{Pekker13}  D. Pekker, G. Refael, E. Altman,  E. Demler, and V. Oganesyan,
 {\it Phys. Rev. X} {\bf4}, 011052 (2014).
 
 \bibitem{Vasseur15}  R. Vasseur, A. C. Potter, and S.A. Parameswaran,
 {\it Phys. Rev. Lett.} {\bf114}, 217201 (2015).
 
  \bibitem{Potter15b} A. C. Potter, R. Vasseur,  and S.A. Parameswaran, arXiv:1501.03501 (2015)

  \bibitem{Agarwal15} K. Agarwal, S. Gopalakrishnan, M. Knap, M. Muller, and E. Demler, {\it Phys. Rev. Lett.} {\bf114}, 160401 (2015).
  
    \bibitem{BarLev15} Y. B. Lev, G. Cohen, and D. R. Reichman, {\it Phys. Rev. Lett.} {\bf114}, 100601 (2015).
  
  
  \bibitem{Gopalakrishnan15} S. Gopalakrishnan, M. Mueller, V. Khemani, M. Knap, E. Demler, D. A. Huse,  arXiv:1502.07712 (2015)



\bibitem{Vosk13} R. Vosk and E. Altman,  {\it Phys. Rev. Lett.} {\bf112}, 217204 (2014).

\bibitem{Serbyn13} M. Serbyn, Z. Papic, and D. A. Abanin,  {\it Phys. Rev. Lett.} {\bf110}, 260601 (2013).



\bibitem{Bauer13} B. Bauer and C. Nayak, {\it J. Stat. Mech.}, P09005 (2013).

\bibitem{Swingle13} B. Swingle, arXiv:1307.0507 (2013).





\bibitem{Schiulaz13} M. Schiulaz, M. Muller, {\it AIP Conf. Proc.} {\bf1610}, 11 (2014).

\bibitem{Khemani15} V. Khemani,	R. Nandkishore, S. L. Sondhi, Nat. Physics {\bf11}, 560 (2015). 


\bibitem{finitedensity} We use temperature interchangeably with finite energy density. Since we are considering isolated quantum systems out-of-equilibrium, temperature is not strictly well-defined.

\bibitem{parity} Since parity is conserved, the strings could just as well be written to lie entirely in the SPT phase.



\bibitem{Haldane83} F. D. M. Haldane, {\it Phys. Rev. Lett.} {\bf50}, 1153 (1983).


\bibitem{Affleck87} I. Affleck, T. Kennedy, E. H. Lieb, H. Tasaki, {\it Phys. Rev. Lett.} {\bf59}, 799 (1987).

\bibitem{Alicea11} J. Alicea, Y. Oreg, G. Refael, F. von Oppen, M. P. A. Fisher, {\it Nat. Phys.} {\bf7}, 412 (2011).

\bibitem{Carmele15} A. Carmele, M. Heyl, C. Kraus, M. Dalmonte,  arXiv:1507.06117 (2015).


\bibitem{Iadecola15} T. Iadecola, L. H. Santos, C. Chamon, arXiv:1503.07871 (2015)


\bibitem{SuppInfo} See Supplemental Material 

\end{thebibliography}
\end{document}


\title{Supplemental Material for Many-body Localization Protected Quantum State Transfer}
\maketitle
\hspace{56mm}N. Y. Yao, C. R. Laumann, A. Vishwanath

\vspace{10mm}

\noindent  {\bf Static coherence times}---Here, we provide additional numerics and analytic details regarding the state transfer protocol. First, we consider the coherence time of the edge qubit in the infinite temperature, clean, interacting system.  Fig.~S1a depicts the trace fidelity as a function of time for a length $L=8$ SPT chain. Fitting the decay envelope, $e^{-t/T_2^*}$, of $\Sigma^x$, $\Sigma^z$ allows one to extract $T_2^*$. Moreover, by varying the interaction strength $V$, while holding all other parameters fixed, we extract the scaling, $T_2^* \sim \lambda/V^2$ (Fig.~S1b). Adding strong disorder brings us to the case of the maintext, where $T_2^* \sim e^L$; an example of this extended coherence is shown in the inset of Fig.~S1a. 

\begin{figure}[h]
	\centering
		\includegraphics[width=4.5in]{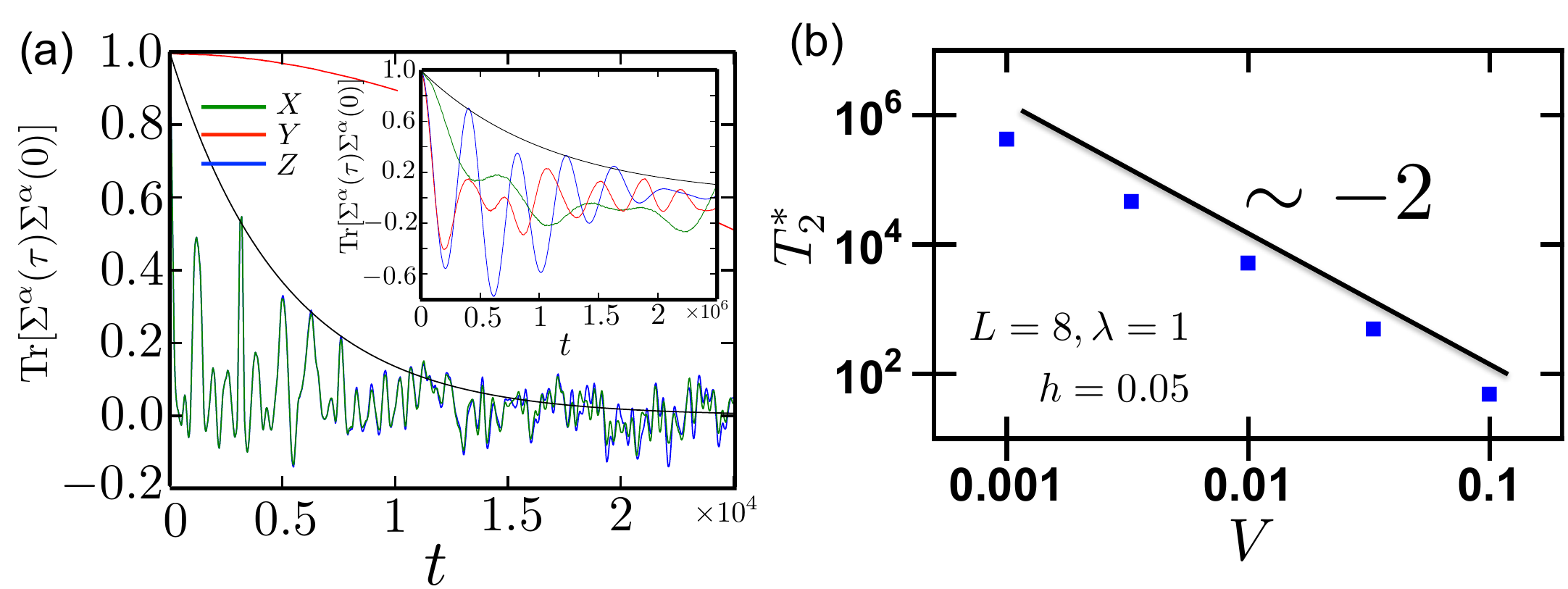}
	\caption{%
	Infinite temperature, clean, interacting case. 
	(a) Quantum information storage in the SPT edge qubit as a function of time with $L=8$, $\lambda =1$, $h=0.05$, $V=0.01$. (inset) Same but for uniform disorder of width $\delta \lambda = 0.7$, $\delta h = 0.05$, $\delta V =0.01$.
	(b) Inverse quadratic scaling of the static coherence time as a function of interaction strength $V$. }
	\label{fig:fig1_v1_schem}
\end{figure}

\noindent {\bf Non-adiabatic transitions}---To further confirm that the dynamic protocol fails in the clean case owing to dispersion of the Wigner string into the bulk, we consider an isolated (non-interacting) SPT chain, as shown in the inset of Fig.~S2. Without the semi-infinite trivial insulator to the left, the edge qubit Pauli operators are simply given by,
\begin{align}
	\label{eq:JQ}
	\Sigma^Z_L &= \sigma^z_0 = \gamma^1_0 \nonumber \\
	\Sigma^Y_L &= \sigma^y_0 \sigma^z_1  = -i \gamma^1_0  \gamma^1_1 \nonumber \\ 
	\Sigma^X_L &= \sigma^x_0 \sigma^z_1  = \gamma^1_1.
\end{align}
To perform an analogous dynamic protocol, we shuttle the edge qubit inwards by ramping down the  first two topological terms $\lambda_0$ (Fig.~S2). We  hold for a time $\tau_W = 100/\lambda$ and then ramp  $\lambda_0$ back to its original value. As in the maintext, at all times during the protocol, there is a unique zero mode with an $\mathcal{O}(1)$ gap ($\sim h$) to the bulk for each chain; thus, if each ramp proceeds slower than this gap scale, the zero mode simply shifts and returns with no phase accumulation. 
Moreover, since the edge operators no longer carry Wigner strings, this is the only relevant adiabaticity condition. 

\vspace{5mm}

\noindent The trace fidelity of this static protocol is shown as a function of the ramp time in Figure~S2. The system size is $L=20$ and $\lambda = 1$, $h=0.05$. One finds that the functional dependence on the ramp time is well characterized by a Landau-Zener-like formula,
\begin{align}
	\label{eq:JQ}
	F \approx 1- e^{\eta h^2 \tau_R}
\end{align}
with $\eta \approx 2.8\pi$. Note that the fidelity of this ``no Wigner string'' dynamic protocol is greater than $95\%$ for all $\tau_R > 150/\lambda$. This is in stark  contrast to Fig.~3a of the maintext where \emph{all} parameters are identical except for the existence of a finite trivial chain to the left of the SPT region. 
The existence of this trivial chain implies the the edge qubit operators carry Wigner strings and thus require a significantly more stringent adiabaticity condition that resolves the bulk mini-gaps. 
\begin{figure}
\center
\includegraphics[width=3.4in]{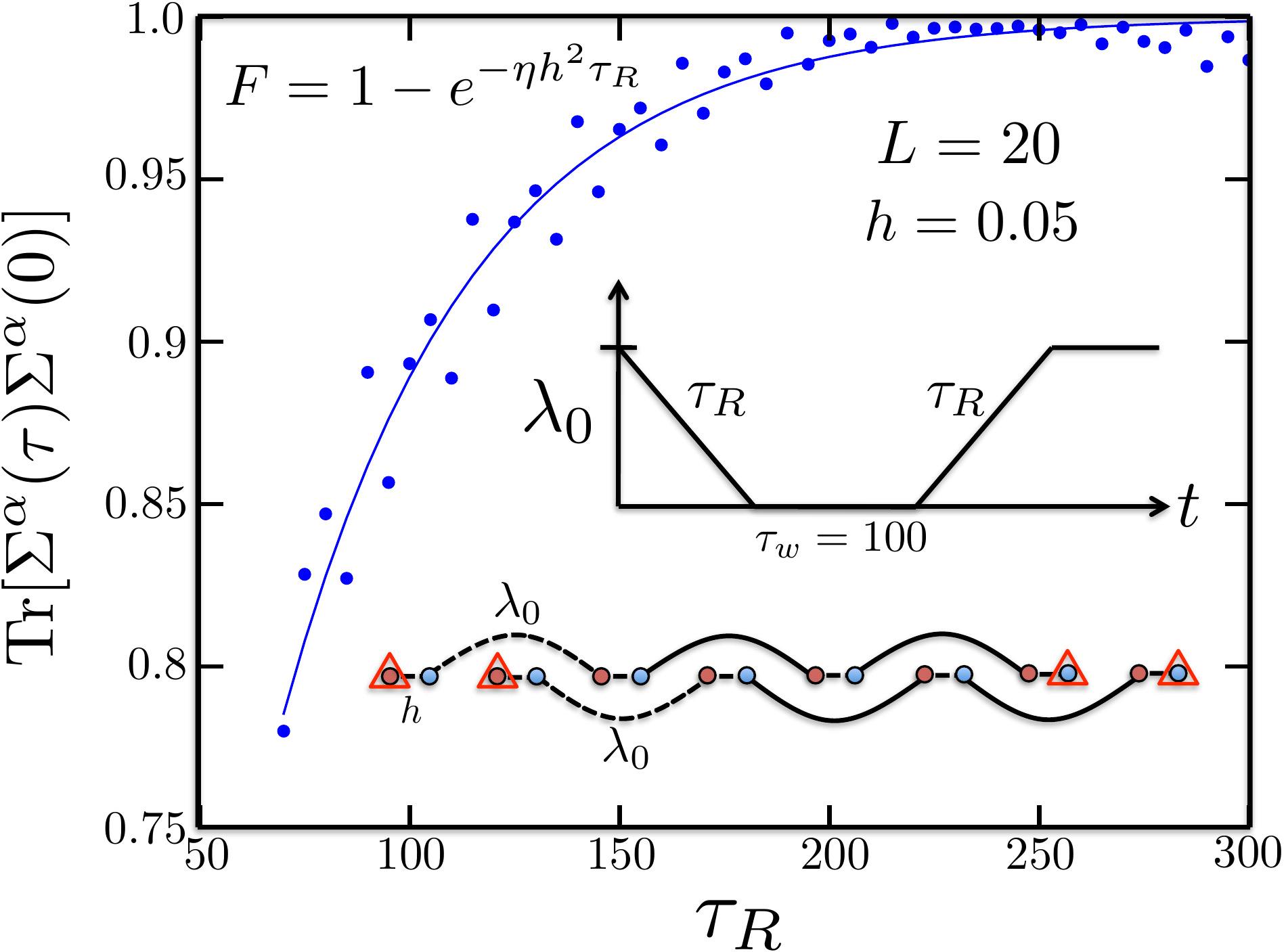}
\caption{Fidelity of the dynamic shuttling protocol for a non-interacting SPT chain as a function of the ramp time. (inset) Pictorial representation of the Majorana chains; note that there is no boundary to a trivial insulator. Also shown is the ramp profile for $\lambda_0$. 
 }
\label{fig:arch1}
\end{figure}

\vspace{5mm}

\noindent {\bf MBL-protected universal two-qubit gate}---We derive the pSWAP enabled universal two-qubit gate between the SPT edge ancillae. As in the maintext, let us assume full control of an edge quantum register that contains the MBL SPT edge qubit and an ancillae qubit. The quantum circuit shown in Fig.~4 of the maintext is calculated for the non-interacting ($V=0$) case by following the Heisenberg evolution of the modes. For simplicity, we neglect the dressing of the edge operators by exponential tails.  
At the end of the circuit, the ancillae operators have transformed as:  $X_L \rightarrow X_L Z_R$,  $Z_L \rightarrow Z_L Z_R$, $X_R \rightarrow -Y_L X_R$, $Z_R \rightarrow Z_R$. Crucially, they are now disentangled from the intermediate SPT chain. 
From the evolution, we can analytically compute the underlying unitary gate \cite{Gottesman99}. Ultimately, this guides the intuition, but the precise gate that is implemented in the interacting case cannot be derived in this fashion; rather, our full numerics demonstrate a universal entangling gate with a finite fidelity reduction for $V \neq 0$.

\vspace{5mm}

\noindent  For the non-interacting case, we explicitly derive the action of the unitary on the basis  $\{ \uparrow \uparrow , \downarrow \uparrow,   \uparrow \downarrow, \downarrow \downarrow   \}$ (of left and right ancillae qubits). Let us start with  $\left |  \uparrow \uparrow  \right\rangle$. This is a $+1$ eigenstate of both $Z_L$ and $Z_R$ and thus, it must also be a $+1$ eigenstate of the final transformed operators $Z_L Z_R$ and $Z_R$. This immediately implies that $U \left |  \uparrow \uparrow  \right\rangle = \alpha \left |  \uparrow \uparrow  \right\rangle $. The other three cases are given by,
\begin{align}
	\label{eq:transform}
	\left | \downarrow \uparrow \right \rangle &= X_L \left |  \uparrow \uparrow  \right\rangle \nonumber \\ 
	U \left | \downarrow \uparrow \right \rangle &= U X_L U^\dagger U \left |  \uparrow \uparrow  \right\rangle\nonumber \\ 
	U \left | \downarrow \uparrow \right \rangle &= X_L Z_R U \left |  \uparrow \uparrow  \right\rangle \nonumber \\
	U \left | \downarrow \uparrow \right \rangle &= \alpha X_L Z_R  \left |  \uparrow \uparrow  \right\rangle \nonumber \\
	U \left | \downarrow \uparrow \right \rangle &= \alpha   \left |  \downarrow \uparrow  \right\rangle 
\end{align}
\begin{align}
	\label{eq:transform}
	\left | \uparrow \downarrow \right \rangle &= X_R \left |  \uparrow \uparrow  \right\rangle \nonumber \\
	U \left | \uparrow \downarrow \right \rangle &= U X_R U^\dagger U \left |  \uparrow \uparrow  \right\rangle \nonumber \\
	U \left | \uparrow \downarrow \right \rangle &= -Y_L X_R U \left |  \uparrow \uparrow  \right\rangle \nonumber \\
	U \left | \uparrow \downarrow \right \rangle &= -i \alpha  \left |  \downarrow \downarrow  \right\rangle 
\end{align}
\begin{align}
	\label{eq:transform}
	\left | \downarrow \downarrow \right \rangle &=  X_L \left |  \uparrow \downarrow  \right\rangle \nonumber \\
	U \left | \downarrow \downarrow \right \rangle &= U X_L U^\dagger U \left |  \uparrow \downarrow  \right\rangle \nonumber \\
	U \left | \downarrow \downarrow \right \rangle &= X_L Z_R U \left |  \uparrow \downarrow  \right\rangle \nonumber \\
	U \left | \downarrow \downarrow \right \rangle &= -i \alpha X_L Z_R  \left |  \downarrow \downarrow  \right\rangle \nonumber \\
	U \left | \downarrow \downarrow \right \rangle &= i \alpha  \left |  \uparrow \downarrow  \right\rangle .
\end{align}
Thus, the final unitary gate between the ancillae qubits is given by,
\begin{align}
U= \left (
\begin{matrix} 
 1 & 0 &0  &0 \\ 
0& 1 & 0 &0 \\
   0 & 0 & 0 & -i  \\
      0 & 0 & i & 0 
\end{matrix} \right).
\end{align}
which is a controlled-$Y$ gate where the control is right ancillae  and the target is the left ancillae qubit. This gate is universal.